\documentstyle[graphics,twocolumn]{mn}
\newcommand{\mincir}{\raise
-2.truept\hbox{\rlap{\hbox{$\sim$}}\raise5.truept 
\hbox{$<$}\ }}
\newcommand{\magcir}{\raise
-2.truept\hbox{\rlap{\hbox{$\sim$}}\raise5.truept
\hbox{$>$}\ }}
\newcommand{\minmag}{\raise-2.truept\hbox{\rlap{\hbox{$<$}}\raise
6.truept\hbox
{$>$}\ }}

\newcommand{\lya}{Lyman-$\alpha$~}

\newcommand{\be}{\begin{equation}}
\newcommand{\ee}{\end{equation}}
\newcommand{\ba}{\begin{eqnarray}}
\newcommand{\ea}{\end{eqnarray}}
\newcommand{\brr}{\begin{array}}
 
\newcommand{\err}{\end{array}}
\newcommand{\bc}{\begin{center}}
\newcommand{\ec}{\end{center}}

\newcommand{\et}{{\it et al.~}}
\newcommand{\ltsima}{\mbox{$\; \buildrel < \over \sim \;$}}
\newcommand{\apj}{\mbox ApJ} \title[Dark energy effects on the
Lyman-$\alpha$ forest] {Dark energy effects on the Lyman-$\alpha$
forest} \author[M. Viel, S. Matarrese, T. Theuns, D. Munshi, Y. Wang]
{M. Viel $^{1, 2}$ S. Matarrese $^{2, 3}$, Tom Theuns $^{4}$, D. Munshi
$^{1,5}$, Yun Wang $^{6}$ \\ $^1$ Institute of Astronomy, Madingley
Road, Cambridge CB3 0HA\\ $^2$ Dipartimento di Fisica `Galileo
Galilei', via Marzolo 8, I-35131 Padova, Italy \\ $^3$ INFN, Sezione di
Padova, via Marzolo 8, I-35131 Padova, Italy \\ $^4$ Institute for
Computational Cosmology, Department of Physics, University of Durham,
South Road, Durham, DH1 3LE\\ $^5$ Astrophysics Group, Cavendish
Laboratory, Madingley Road, Cambridge CB3 0HE\\ $^6$ Department of
Physics and Astronomy, University of Oklahoma, Norman, OK73019, USA\\
\\}

\begin{document}

\maketitle

\begin{abstract}
In quintessence models, the dark energy content of the universe is 
described by a slowly rolling scalar field whose pressure and energy 
density obey an equation of state of the form $p=w \rho$;  $w$ is
in general a function of time such that $w<-1/3$, in order to drive the
observed acceleration of the Universe today. The cosmological constant
model ($\Lambda$CDM) corresponds to the limiting case $w=-1$. In this
paper, we explore the prospects of using the Lyman-$\alpha$ forest to
constrain $w$, using semi-analytical techniques to model the
intergalactic medium (IGM).  A different value of $w$ changes both the
growth factor and the Hubble parameter as a function of time. The
resulting change in the optical depth distribution affects the optical
depth power spectrum, the number of regions of high transmission per
unit redshift and the cross-correlation coefficient of spectra of
quasar pairs. These can be detected in current data, provided we have
independent estimates of the thermal state of the IGM, its ionization
parameter and the baryon density.
\end{abstract}


\section{Introduction}
\label{intro}
Observed cosmic microwave background anisotropies convincingly
demonstrate that the universe is spatially flat (de Bernardis \et 2002;
Netterfield \et 2002; Pryke \et 2002). The luminosity-distance
relation, as determined from high redshift type Ia Supernovae
(Garnavich \et 1998; Riess \et 1998; Perlmutter \et 1999), requires a
spatially flat universe to be currently dominated by some type of dark
energy, which is nearly homogeneous and has negative pressure, such as
for example a cosmological constant. Several independent lines of
argument also favour a low-density, vacuum-energy dominated universe,
for example the abundance of clusters (Bahcall \et 2003) and their
X-ray properties (Allen \et 2002), the clustering of galaxies
(Efstathiou \et 2002; Verde \et 2002), estimates of the age of the
universe and the current value of the Hubble parameter.

The energy density associated with the cosmological constant may
actually decay in time. Such quintessence models (e.g. Caldwell \et
1998; henceforth QCDM models) have an equation of state $p=w\, \rho$,
with $w<-1/3$, where $w=-1$ corresponds to the more familiar
cosmological constant model. Recently, a number of observational tests
have been proposed to measure $w$ and its redshift dependence (Kujat
\et 2001; Wang \& Garnavich 2001; Matsubara \& Szalay 2002; Gerke \&
Efstathiou 2002).

In this {\it Letter} we focus on the prospects of using the \lya forest
to constrain $w$. The \lya forest region in quasar (QSO) spectra
contains hundreds of hydrogen \lya absorption lines (see Rauch 1998 for
a review), most of which are produced by small density fluctuations in
the intervening intergalactic medium (Cen \et 1994; Bi \& Davidsen
1997, hereafter BD97). Since these structures are still reasonably
linear, they are good probes of the underlying large-scale matter
distribution, and hence are sensitive to the cosmological model. For
example, Hui \et (1999) and McDonald (2001) suggested to apply the 
Alcock-Paczinski test (Alcock \& Paczinski 1979) to estimate the cosmological
constant and Croft \et (2002) constrained the dark matter power
spectrum on large scales. The \lya forest is promising in that it can
be used over a larger redshift range than most other potential probes
of $w$.

Here we use semi-analytical models of the \lya forest, which have been
shown to be successful in reproducing reasonably well most of its
observed properties, such as the column density distribution function,
the line-widths distribution and the number of lines per unit redshift
(BD97; Viel \et 2002a, 2002b; Roy Choudury \et 2001).  Hydrodynamical
simulations are required for more accurate predictions on smaller
scales (e.g. Theuns \et 1998, 2002; Dav\'e \et 1999; Bryan \et
1999). We examine several statistics of the flux spectrum, such as the
probability distribution function of the optical depth, the optical
depth power spectrum, the number of underdense regions per unit
redshift and the cross-correlation coefficient of spectra of quasar
pairs. The first two statistics are not directly observable but they
can be recovered from the flux distribution by looking at higher order
lines through pixel optical depths techniques (Sect. 3).
Section~2 briefly reviews the properties of quintessence models and
describes our simulation techniques. Section~3 contains our results
and a brief discussion.

\section{Simulations of \lya spectra in quintessence models}
The quintessence parameter $w$ influences the shape of the linear
matter power spectrum $P(k,z) \propto k^n\, {\cal T}^2(k,z)\, D^2(z)$,
the linear growth factor of density perturbations $D(z)$ and the Hubble
parameter $H(z)$. Here $k$ is the wavenumber, ${\cal T}$ the transfer
function and $n$ the spectral index. Useful expressions for the growth
factor of density perturbations and of linear peculiar velocities can
be found in Lahav {\it et al.} (1991), for $\Lambda$CDM, and in Wang \&
Steinhardt (1998), for QCDM. The Hubble parameter at redshift $z$ is
$H(z)=H_0
\left[\Omega_{m}(1+z)^3+\Omega_{k}(1+z)^2+\Omega_{DE}(1+z)^{3+3w}\right]^{1/2}$,
with $H_0=100\,h$\,km s$^{-1}$ the present-day Hubble parameter and
$\Omega_{k} = 1-\Omega_{m}-\Omega_{DE}$.  $\Omega_{DE}$ represents the
energy density of the cosmological constant ($\Omega_{\Lambda}$) or the
quintessence ($\Omega_Q$), in units of the critical density.
 
In Figure~1 we plot ${\cal T}(k,z)$, $D(z)$ and the Hubble parameter $H(z)$,
for several values of $w$, here assumed to be constant in time, for
simplicity, using the fits of Ma \et (1999).  If we compare the
transfer function of QCDM and $\Lambda$CDM at $z=2$ one can see that
there are differences only on very large scales, $k\ltsima 0.01\, h$\,
Mpc$^{-1}$. The Hubble parameter in QCDM models with $w=-0.8$, $-0.6$ and
$-0.4$ differs by 6, 14 and 25 per cent at redshifts $z\sim 0.75$, $z\sim
1$ and $z\sim 1.2$ from those of the $\Lambda$CDM model (panel b). For
those values of $w$, the growth factors $D(z)$ (panel c) at $z\sim 2$
are larger than for the $\Lambda$CDM model by 5, 15 and 30 per cent
respectively.

\begin{figure*}
\center\resizebox{.7\textwidth}{!}{\includegraphics{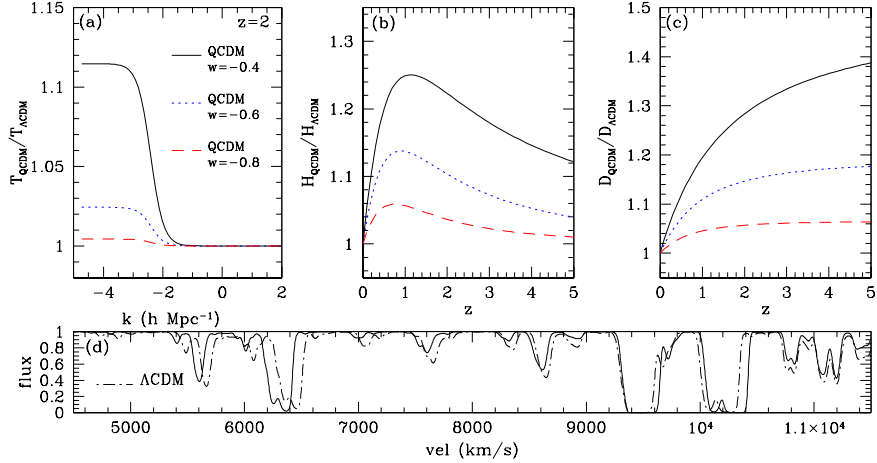}}
\caption{{\protect\footnotesize{Panel a): ratio of transfer functions ${\cal T}_{QCDM}/{\cal
T}_{\Lambda CDM}$ at z=2. Panel b): ratio of Hubble parameters
$H_{QCDM}/H_{\Lambda CDM}$ as a function of redshift. Panel c): ratio
of growth factors $D_{QCDM}/D_{\Lambda CDM}$ as a function of
redshift. The three curves shown are for three quintessence models with
$w=-0.4$ (continuous line), $w=-0.6$ (dotted line) and $w=-0.8$ (dashed
line). Panel d): example of simulated spectra, it is shown $\Lambda$CDM
spectra (dot-dashed line) and the QCDM with $w=-0.4$ (continuous
line).}}}
\label{p1}
\end{figure*}

The semi-analytical model we use to generate \lya forest spectra is
based on the approximations introduced by Bi and collaborators (BD97;
see Viel \et (2002a) for more details). The basic assumption is that
the low-column density Ly$\alpha$ forest is produced by smooth
fluctuations in the intergalactic medium (IGM), which arise as a result
of gravitational growth of perturbations (e.g. Schaye 2001). Briefly,
we start by generating correlated Gaussian random fields to represent
the linear density and peculiar velocity dark matter fields along a
sight-line, for a given linear matter power spectrum. The
linear density perturbations in the IGM $\delta_0^{\rm IGM}({\bf x},
z)$ are related to those in the underlying dark matter by a
convolution, $ \delta_0^{\rm IGM} ({\bf k}, z) = W_{\rm IGM}(k,z)
\delta_0^{\rm DM}({\bf k},z)$. For the smoothing kernel $W$ we use the
Gaussian filter $W_{IGM}=\exp(-k/k_f)^2$ (Gnedin \et
2002), which is a good approximation to model the effects of gas
pressure in the linear regime. The smoothing scale $k_f$ is related to
the Jeans length in a way which depends on the thermal history of the
IGM (Gnedin \& Hui 1998; Matarrese \& Mohayaee 2002).

To account for the fact that the IGM which produces most of the \lya
forest is actually mildly non-linear, we follow BD97 and adopt a
simple lognormal model (Coles \& Jones 1991) for the IGM local
density. This is a strong assumption but we think that at these
redshift and for the large scale properties investigated here the
log-normal model can be a good approximation to the more accurate
hydro-dynamical simulations.  We assume the gas to be in
photoionization equilibrium with an imposed uniform UV-background,
which we scale so that the mock spectra have the same effective
optical depth $\tau_{\rm eff}$ as observed (see Kim \et 2002 and
Bernardi \et 2002, for recent determinations of $\tau_{eff}(z)$). We
use the values of Schaye \et (2000) for the normalization $T_0$ and
slope $\gamma$ of the temperature-density relation,
$T=T_0(\rho/\langle\rho\rangle)^{\gamma-1}$, in the IGM.  Finally, we
artificially broaden the spectra to a resolution of FWHM = 6.6 km
s$^{-1}$ to mimic high quality VLT/UVES or Keck/HIRES spectra, and add
Gaussian noise with signal to noise of 50 (see e.g. Theuns \et 2000
for more details). Spectra generated with this procedure have been
shown to produce flux probability and line-width distribution
functions in reasonable agreement with observations (BD97).

We will compare the \lya forest in a $\Lambda$CDM cosmological model
with three different QCDM models, with $w=-0.4,-0.6,-0.8$,
respectively. For all of these models, we assume that the cosmological
parameters are $\sigma_8=0.7,\Omega_{m}=0.3, h=0.67,
\Omega_b\,h^2=0.020, \, \Omega_{DE}=0.7$. We simulate 10 different
randomly generated spectra, with the same set of random phases for
each spectrum, in the redshift range $1.8<z<2.2$ for each model,
assuming $\tau_{eff}=0.15$ (Kim \et 2002), $\gamma=1.3$ and
$T_0=10^{4.1}$ K. The filtering scale is $k_f\sim 1.7\, k_J$ (with
$k_J$ the Jeans wave-number), which is a reasonable estimate if HI
reionization takes place at $z\sim 7$ (Gnedin \& Hui 1998). The
spectra are approximately $40,000$ km s$^{-1}$ long. In addition, we
simulate spectra of QSO pairs with a given angular separation in the
range 10--90 arcsec, using the procedure described in Viel \et
(2002a). The cross-correlation coefficient spectra of pairs is
estimated using the definition of Viel \et (2002a) from a set of 8
pairs for each separation.  In panel d) of Figure 1 we show, for a
qualitative comparison, a chunk of two simulated spectra for the
$\Lambda$CDM and QCDM, with $w=-0.4$, model. We have decided to focus
our analysis around $z\sim 2$ in view of a comparison with
observations which will be made in a future work.

\section{Results}

In this section we will study the effects of dark energy on the
simulated spectra.  The differences between the QCDM and $\Lambda$CDM
model, which could be tested through \lya absorptions, are the
following: the linear dark matter power spectrum $P(k,z)$; the evolution
of the cosmological parameters and in particular of the Hubble
parameter $H(z)$; the growth factors $D(z)$ of the linear density
perturbations.  

The simulated \lya flux power spectrum agrees
reasonably well with the linear dark matter power spectrum at large
scales, while on smaller scales non-linear effects, thermal broadening
and noise can produce differences. This is roughly true even if we do
not correct for redshift-space distortions and compare directly the
flux power spectrum of simulated absorption spectra and the linear dark
matter power spectrum. From Figure \ref{p1} (panel a) we can see that
the QCDM transfer function differs from the $\Lambda$CDM one only on
scales larger than 100 comoving Mpc.  Such scales are comparable to the
extent of the \lya forest region in a QSO spectrum, and hence this
signature of $w\ne -1$ would be difficult to detect since it would
require a large number of spectra to estimate the power on these
scales. More importantly, in high resolution spectra every attempt to
recover the power spectrum on scales larger than 10 comoving Mpc is
challenging due to the continuum fitting uncertainties (Croft \et
1999).

A second difference between QCDM and $\Lambda$CDM is the different
evolution of the cosmological parameters, such as $H(z)$.  We have
explored the effect of a different Hubble parameter and different
growth factors values at redshift $z=2$ on the simulated \lya optical
depth, by fixing the remaining parameters. The effect is clear: a
larger Hubble parameter produces in general a broader optical depth
distribution with the peak shifted to lower values. In fact, with a
higher value of the Hubble parameter, the same real-space absorber
intersected by the sight-line will affect a larger region of the
redshift-space spectrum.

In the same way, we explore the effect of different growth factors. If
the growth factor is increased, the (non-linear) log-normal mapping
acts in such a way that typically the low-density regions become less
dense and the high-density regions more dense.  Thus, with a larger
growth factor, as the one predicted for QCDM models compared to
$\Lambda$CDM ones (Figure \ref{p1}, panel c), the simulated optical
depth shifts to smaller values and its probability distribution
function (pdf) becomes broader. This effect can be seen also in the
simulated flux in panel d) of Figure \ref{p1}, for example at $\sim
6000$ km s$^{-1}$ this is particularly evident.

Our first conclusion is that a larger growth factor produces the same
effect of a higher value of the Hubble parameter: a shift of the
simulated \lya optical depth to smaller values and a broader
distribution, if the other parameters remain fixed.

\begin{figure*}
\center
\resizebox{0.72\textwidth}{!}{\includegraphics{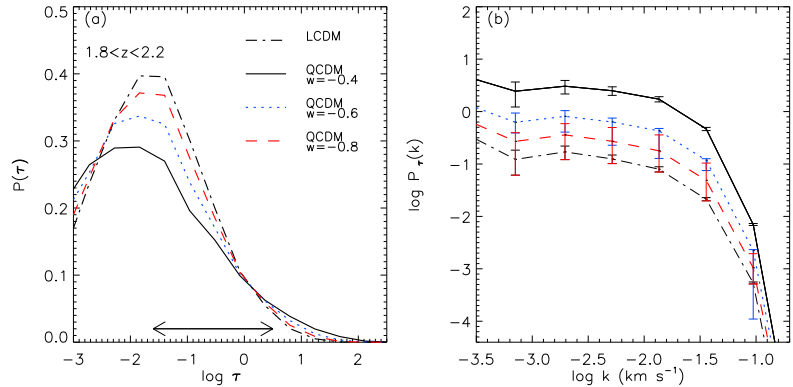}} 
\caption{{\protect\footnotesize{Panel a): probability distribution
function of the optical depth (panel a) for QCDM
model with $w=-0.4$ (continuous line), $w=-0.6$ (dotted line),
$w=-0.8$ (dashed line) and $\Lambda$CDM model (dot-dashed line). The
interval in which the optical depth can be recovered directly from the
flux is represented by the double arrow (a larger range can be used with
pixel optical depth technique, see text).  Panel b): power spectrum of
the optical depth; error bars represent the error of the mean.}}}

\label{fig2}
\end{figure*}

To see these effects combined together, we plot in panel a) of Figure
\ref{fig2} the probability distribution function of the simulated
optical depth for $\Lambda$CDM (dot-dashed line), QCDM with $w=-0.4$
(continuous line), $w=-0.6$ (dotted line) and $w=-0.8$ (dashed
line). We choose to plot the probability distribution function also for
$\tau$ values smaller than those which can actually be recovered by
standard techniques to better appreciate the differences between
cosmological models. We can see that the simulated optical depth
pdf results in different distributions
for the four cosmological models. Since the flux is the observed
quantity, the inferred optical depth can be trusted only in the range
$-1.6\mincir \log(\tau)\mincir 0.5$ (corresponding to fluxes in the
range 0.04-0.97). With a more sophisticated analysis, based on pixel
optical depth techniques, this range can be considerably extended
(Aguirre \et 2002 and references therein).  In panel b) of Figure
\ref{fig2} we plot the optical depth power spectrum for the QCDM and
the $\Lambda$CDM model, with Poissonian error bars. We can see that on
scales $\log k \,(\rm{km\,s}^{-1}) <-1.5$ the four cosmological models
are distinguishable: while the shape is very similar, the normalization
is different even if we choose the same value for $\tau_{eff}$. This is
the effect of different amounts of non linearity in the four
cosmological models, which influences the normalization of the curves.

\begin{figure*}
\center
\resizebox{0.72\textwidth}{!}{\includegraphics{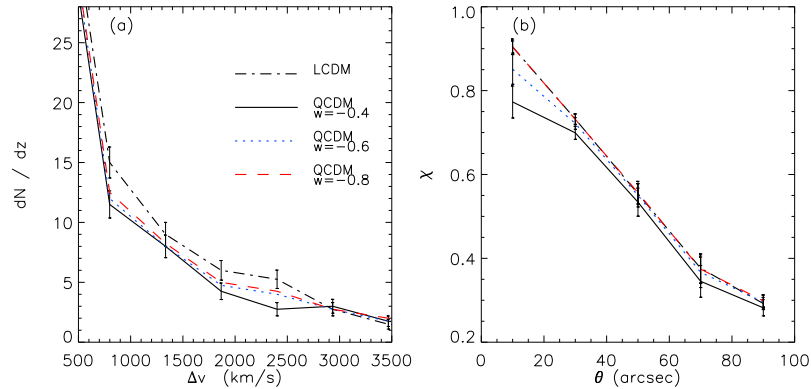}} 
\caption{{\protect\footnotesize{Number of underdense regions per unit
redshift as a function of void size in km s$^{-1}$ (panel a) for the
three QCDM models with $w=-0.4$ (continuous line), $w=-0.6$ (dotted
line), $w=-0.8$ (dashed line) and $\Lambda$CDM model (dot-dashed
line). The simulated sample consists of 10 spectra, spanning the
redshift range $1.8<z<2.2$ for each cosmological model.
Cross-correlation coefficient, $\chi$, as a function of QSO pair
separation in arcsec (panel b). The simulated sample consists of 8 QSO
pairs for each separation shown. Poissonian error bars in panel a),
shown only for two different models for clarity. In panel b) the error
bars are the errors of the mean value.}}}
\label{fig3}
\end{figure*}

In Figure \ref{fig3} we plot two more observational quantities: the
number of underdense regions per unit redshift (panel a) and the
cross-correlation coefficient $\chi$ estimated from the spectra of QSO
pairs at different separations (panel b).  The definition of underdense
region adopted here is the following: if we are searching for an
underdense region of size $\Delta v$, we preliminarily smooth the
simulated spectra on this scale (top-hat smoothing) and then we define
as an underdensity a region for which the flux is larger than the mean
flux at that redshift, estimated through $\tau_{eff}$.
One can see that the predicted number of underdense regions per unit
redshift is different for the four cosmological models. The number of underdense regions of intermediate sizes, between 700 km
s$^{-1}$ and 2700 km s$^{-1}$, which correspond to sizes in the range
10-40 comoving Mpc in a $\Lambda$CDM universe, is larger in the
$\Lambda$CDM model than in the QCDM ones.  This is due to the combined
effects of a different Hubble parameters and a different growth factor
for density perturbations, as we saw in panel a) of Figure \ref{fig2},
which produce more rarefied underdense region and denser high density
regions in QCDM models than in $\Lambda$CDM models. In particular, we
have found that the number of underdense regions of size 2500 km
s$^{-1}$ for $\Lambda$CDM can be a factor of 2 and 1.5 larger than in
QCDM model with $w=-0.4$ and $w=-0.8$ at $z=2$, respectively.

The cross-correlation coefficient $\chi$ is plotted in panel b) of
Figure \ref{fig3}, this number quantifies the amount of `coherence' in
the transverse direction between spectra of QSO pairs. The information
contained in the transverse direction could be powerful and has already
been analysed in the past (Miralda-Escud\'e \et 1996; Viel \et 2002a).
This test is in a sense similar to the more sophisticated
Alcock-Paczinski test (Hui \et 1999; McDonald \& Miralda-Escud\'e 1999;
McDonald 2001) and aims at discriminating between different
cosmological models. Here, we have simulated 8 QSO pairs at $z \sim 2$
for five separations (10, 30, 50, 70 and 90 arcsec) and computed the
mean value and the error of the mean for this coefficient at each
separation. The four models are very similar and the error bars
overlap. Only at angular scales $\mincir 20$ arcsec there are some
differences between the models. However, the number of
pairs considered here is marginally sufficient to discriminate between a
$\Lambda$CDM model and a QCDM with $w=-0.4$. We have estimated that 25
pairs with these small separations are needed to distinguish between
the $\Lambda$CDM model and a QCDM with $w=-0.6$ at a 3$\sigma$ level.

Dark energy in the form of quintessence in the \lya forest seems to be
not easily distinguishable from a cosmological constant. Even if the
thermal state of the IGM is completely known, the uncertainties on the
values of the baryon fraction (Burles \et 2001) and especially on the
ionization parameter (Scott \et 2000), are so large that it would be
difficult to recover the evolution of $H(z)$. We have found that a
QCDM model with $w=-0.4$ requires an ionization parameter a factor of
3 lower than in a $\Lambda$CDM model (while for $w=-0.8$ it is a
factor of 1.25 lower), in order to match observed
$\tau_{eff}$. Uncertainties on the value of $\tau_{eff}$ are of
the order of 40\%, but in the near future, with larger data set
available, this uncertainty will become significantly smaller. In
principle, an accurate determination of $\tau_{eff}$, with a precision
of $10-20\%$ could already be used to discriminate between QCDM with
$w=-0.4$ and $\Lambda$CDM models, since the present uncertainties on
the ionization parameter are of the order of 50\%. Statistics based on
the optical depth like the pdf or the power spectrum, instead of the
flux, can be much more useful in discriminating between different
cosmological models. This is due to the fact that the optical depth is
very sensitive to the amount of non-linearity introduced in the
different cosmological models. However, in order to do that, it will
be necessary to use pixel optical depth techniques, using higher order
Lyman lines, to get reliable estimates of the optical depths both in
saturated and in high transmissivity regions.

{\it In summary}, we have presented a preliminary analysis of simulated
spectra in QCDM models and in a $\Lambda$CDM model based on
semi-analytical models. The main differences can be found in the
different evolution of $H(z)$ and of the linear growth factor of
density perturbations $D(z)$. Both these effects act in a similar
way. If we fix the remaining parameters and compare the probability
distribution function of \lya optical depth between QCDM and
$\Lambda$CDM we have found that the distribution in the QCDM model is
broader than in the $\Lambda$CDM case. This means that in general high
transmission regions are more rarefied and high density regions are
denser in QCDM than in a $\Lambda$CDM universe.  The optical depth
power spectrum is different on intermediate and large scales, even if
the normalization $\tau_{eff}$ is the same, because of the different amounts
of non linearity predicted for the models. Detecting the underdense
regions could be promising. In fact, we have found that the number of
intermediate size underdense regions, with sizes of the order of 10
comoving Mpc, is larger in a $\Lambda$CDM model than in QCDM ones.
Another statistics we have proposed concerns the the cross-correlation
coefficient between the spectra of QSO pairs. However, in this case,
the differences can be appreciated only for separations smaller than 20
arcsec and the number of pairs should be larger than 20 to distinguish
between QCDM and $\Lambda$CDM cosmological models at a 3$\sigma$ level.

%\end{figure*}

\section*{Acknowledgments}
We thank T.-S. Kim and M. Haehnelt for useful discussions. This work
was partially supported by the European Community Research and Training
Network `The Physics of the Intergalactic Medium' and NSF CAREER grant
AST-0094335. DM was supported by PPARC grant RG28936. TT thanks PPARC
for the award of an Advanced Fellowship.

\end{document}